\documentclass[11pt]{article}

\usepackage{a4wide,amssymb}
\newtheorem{theorem}{Theorem}[section]    
\newcommand{\qed}{\hfill{$\rule{6pt}{6pt}$}} 

\newcommand{\ket}[1]{| #1 \rangle}
\newcommand{\bra}[1]{\langle #1 |}

\newcommand{\defeq}{\stackrel{\Delta}{=}}

\newcommand{\cP}{{\cal P}}

\newcommand{\cA}{{\cal A}}

\newcommand{\np}{\mbox{NP}}

\newcommand{\xor}{{\oplus}}

\newcommand{\Vpcp}{V_{\mbox{\scriptsize PCP}}}

\title{Entanglement-Resistant Two-Prover Interactive Proof Systems 
and Non-Adaptive Private Information Retrieval Systems
} 
\author{
Richard Cleve \thanks{Email: {\sf cleve@cs.uwaterloo.ca}.}
\and
Dmitry Gavinsky \thanks{Email: {\sf dmitry.gavinsky@gmail.com}.}
\and
Rahul Jain 
\thanks{Email: {\sf rjain@cs.uwaterloo.ca}.} \\
Computer Science Department and Institute for Quantum Computing
\thanks{University of Waterloo, 200 University Ave.\ West, Waterloo, ON N2L 3G1,
Canada.}
}

\date{}

\begin{document}
\maketitle
\begin{abstract}
\noindent
We show that, for any language in $\np$, there is an entanglement-resistant 
constant-bit two-prover interactive proof system with a constant 
completeness vs.\ soundness gap.
The previously proposed \textit{classical} two-prover constant-bit 
interactive proof systems are known not to be entanglement-resistant.
This is currently the strongest expressive power of 
any known constant-bit answer multi-prover interactive proof system 
that achieves a constant gap.
Our result is based on an ``oracularizing" property of certain private 
information retrieval systems, which may be of independent interest.
\end{abstract}

\section{Introduction}

Properties of interactive proof systems have been shown to change in 
fundamental ways when the underlying setting changes from classical 
information to quantum information.
The first result along these lines was discovered by Watrous 
\cite{Watrous99}, and several subsequent results have occurred.

In the present paper, we are concerned with \textit{multiprover 
interactive proof systems} (MIPs), first proposed (in the classical 
setting) by Ben-Or \textit{et al.}~\cite{Ben-OrG+88}. An example of 
such a system, for $\mathsf{3SAT}$, is where the first prover is sent a clause 
of the formula and the second prover is sent a variable from the 
clause. The first prover must give a partial truth assignment that 
satisfies the clause and the second prover must give an assignment 
to the variable that is consistent with the first prover's (this 
protocol occurs in several places in the literature, e.g., 
\cite{Hastad01}). Classically, the completeness probability of this 
system is 1, whereas the soundness probability is at most $1 - 
\frac{1}{3m}$, where $m$ is the number of clauses. Roughly speaking, 
the role of the second prover is to ``oracularize" the first prover: 
to make it respond to queries as an oracle would; if the first 
prover behaves adaptively to queries, this introduces a positive 
probability that the results between the two provers are not 
consistent. In the quantum setting, if the provers are allowed to 
share \textit{a priori} entanglement they can cheat the protocol in 
the sense that there are \textit{unsatisfiable} 3CNF formulas where 
the soundness probability of the protocol is 1 (hence the gap is 
zero) \cite{CleveH+04a}. Thus this particular oracularization 
technique fails in the setting of quantum information (other 
examples are also given in \cite{CleveH+04a}). Note that, in the 
above, quantum information enters the picture by the entanglement 
between the provers; the verifier and its communication with the 
provers remains classical. Interesting results have also been 
obtained for multiprover interactive proof systems where the 
communication between the verifier and provers is quantum.

A major question is how the expressive power of two-prover interactive proof 
systems changes when the provers possess entanglement.
Without entanglement, this is known to be NEXP 
\cite{BabaiF+91,FortnowR+94}.
Since entanglement can potentially increase both the completeness and 
soundness probability, it is not even clear whether the expressive 
power is a subset or superset of NEXP.
In \cite{CleveH+04a} (based on results in \cite{Hastad01}), it is shown 
that a restricted class of MIPs (called $\oplus$-MIPs or XOR-MIPs) has 
the property that, classically, their expressive power equals NEXP; 
whereas, with entangled provers, its expressive power reduces to a 
subset of EXP (see \cite{Wehner06} for a refinement of this result).
Thus, for $\oplus$-MIPs, entanglement strictly reduces their expressive 
power (unless EXP = NEXP).
An $\oplus$-MIP has the simple form where the verifier makes one 
polynomial-length query to each prover, and each prover returns a 
single bit answer to the verifier.
The verifier's acceptance condition is a function of the XOR of the 
two answer bits and the questions.

Our main result is the introduction of a new technique for 
oracularizing provers, based on properties of certain 
\textit{private information retrieval systems }(PIRs). A PIR is a 
system that enables information to be obtained from a database 
without revealing to the database server(s) what the information is 
that is being queried. The framework is two (or more) isolated 
servers who each have a copy of the database, but who cannot 
communicate with each other. Instead of asking an individual server 
for the information (which would reveal the query), each server is 
asked for information and the responses are combined to produce the 
answer. There are ways of doing this such that no individual server 
acquires any information about the actual query being made.

Intuitively, this seems like a natural approach to oracularizing 
provers in a MIP: if the servers have no idea of that is being queried 
in the first place, how can they make their answers adaptive? 
 Although this sounds intuitively compelling, the 
non-adaptiveness property is operationally different from the PIR 
property. This distinction is reminiscent of the distinction 
between malleable cryptography and non-malleable cryptography 
\cite{DolevD+00}. For example, a cryptosystem may be secure in the 
sense that it is not possible to deduce $x$ from an encryption of 
$x$, nevertheless it may be possible from this encryption to 
construct an encryption of $y$ that is somehow related to $x$.

We show that certain PIRs are in fact non-adaptive in the sense that,
not only do they reveal no information about the items queried to the
servers, but they satisfy the additional property that the servers
cannot conspire to make their answers satisfy a property that
non-trivially depends on the queries made. Remarkably, this property is
robust even against
\textit{quantum} servers who have the resource of \textit{a priori} 
entanglement.

Based on this, we show that $\oplus$-MIP has expressive power at least 
that of $\np$ for entangled provers.
This is the strongest expressive power of any known 
constant-bit answer MIP that achieves a constant gap.

Related work is \cite{KempeVidick06,KKMTV07}, where other novel techniques 
are introduced.
In this work, the complexity classes are supersets of $\np$; however, 
the gaps between completeness and soundness probability are smaller and 
the communication from the provers is larger.
Hence these results are incomparable with ours.

\section{Some notation}
We use the following notation in this paper. For $s, t\in \{0,1\}^m$,
let $s \cdot t \in \{0,1\}$ denote the inner product modulo 2 of $s$
and $t$, and $s \oplus t \in \{0,1\}^m$ denote the bitwise
exclusive-or of $s$ and $t$.
For $j \in \{1,2,\dots,m\}$, let $e_j \in \{0,1\}^m$ denote the 
characteristic vector of $\{j\}$, which is 1 in component $j$ and 0 in all other 
components.

\section{Our main result}
Our main result is as follows.
\begin{theorem}
For all $\varepsilon > 0$, for any language $L$ in $\np$, there exists
a two-prover protocol $\cP$ with entangled provers 
of the following form. Let $x \in \{0,1\}^n$ ($n$ large enough
depending on $\varepsilon$) be the input
received by the provers Alice and Bob and the Verifier $V$.
\begin{enumerate}
\item
$V$ generates messages $s$ and $t$, each of length polynomial in $n$,
and a private bit $\delta$; $(s,t,\delta)$ chosen from a certain
polynomial time samplable joint distribution. $V$ then sends $s$,
$t$ to Alice and Bob respectively.
\item
Alice and Bob respond with bits $a$ and $b$ respectively.
\item
$V$ accepts $x$ if and only if $a \oplus b = f_x(s \oplus t, \delta)$, 
where $f_x$ is computable in time polynomial in $n$.
\end{enumerate}

The protocol satisfies the following soundness/completeness properties:
\begin{description}
\item[\bf Completeness:]
If $x \in L$ then there exists a strategy for provers Alice and Bob 
such that $V$ accepts with probability $\geq 1 - 
\varepsilon$. 
\item[\bf Soundness:]
If $x \notin L$ then, for all strategies of provers Alice and Bob, 
$V$ accepts with probability $\leq \frac{1}{2} + 
\varepsilon$.
\end{description}

\end{theorem}

The above theorem immediately implies $\np \subseteq
\oplus$-MIP*[2,1], where $\oplus$-MIP*[2,1] represents that class of
languages acceptable by two entangled prover proof systems with a
single round of interaction and in which verifier only uses the xor of
the bits answered by the provers to decide. 
\section{The PCP system}

Let $L \in \np$ and $\varepsilon > 0$. From~\cite{Hastad01}, there exists a 
probabilistically checkable proof (PCP) system for $L$ of the following 
form. There is a proof verification procedure $\Vpcp$ that, for any 
$n$-bit string $x$ ($n$ large enough
depending on $\varepsilon$), takes an $m$-bit string $w$ as input (where $m 
\in n^{O(1)}$) and accepts or rejects $w$ as a certificate of $x \in 
L$ based on the parity of three bits of $w$ as follows. $\Vpcp$
probabilistically generates distinct $i, j, k \in \{1,2,\dots,m\}$ and $\delta
\in \{0,1\}$, from a certain polynomial time (in $n$) samplable joint
distribution, and accepts if and only if $w_i \oplus w_j
\oplus w_k = f_x(i, j, k, \delta)$, where $f_x$ is a polynomially 
computable function.  The completeness/soundness properties
of the proof system are as follows.
\begin{description}
\item[Completeness:]
For all $x \in L$, there exists a witness string $w \in \{0,1\}^m$ 
such that $\Vpcp$ accepts with probability at least $1 - \varepsilon$.
\item[Soundness:]
For all $x \not\in L$, for all $w \in \{0,1\}^m$, $\Vpcp$ accepts 
with probability at most $\frac{1}{2} + \varepsilon$.
\end{description}

\section{Our protocol via the PIR reduction}

Using the PCP procedure $\Vpcp$ we obtain our protocol $\cP$ as
follows. On receiving input $x$, $V$ interacts with provers Alice and Bob as follows.
\begin{enumerate}
\item
$V$ simulates $\Vpcp$ in the generation of $i, j, k \in \{1,2,\dots,m\}$ 
and $\delta \in \{0,1\}$.
\item
$V$ chooses $s \in \{0,1\}^m$, uniformly distributed and
independently of $i,j,k, \delta$, and set $t  = s \oplus e_i \oplus e_j \oplus e_k$.
\item
$V$ sends $s$ to Alice and $t$ to Bob, receiving one-bit answers $a$
and $b$ from them respectively.
\item
$V$ accepts if and only if $a \oplus b = f_x(i,j,k, \delta)$.
\end{enumerate}

\subsection{Completeness}

If $x \in L$ then we know from the PCP procedure that there exists a
PCP-witness $w \in \{0,1\}^m$.  Consider the strategy in which Alice,
on input $s$ outputs $a = w \cdot s$, and Bob, on input $t$, outputs
$b = w \cdot t$.  Now,
\begin{eqnarray*}
a \oplus b & = & w \cdot (s \oplus t) \\
& = & w \cdot (e_i \oplus e_j \oplus e_k) \\
& = & w_i \oplus w_j \oplus w_k.
\end{eqnarray*}
Therefore  $V$ accepts whenever $\Vpcp$ accepts the PCP string
$w$, and hence the probability of acceptance of $V$ is at least $1 - \varepsilon$.

\subsection{Soundness}

The proof of soundness employs a result about certain XOR games that 
are similar to those analyzed by Linden \textit{et 
al.}~\cite{LPSW06}. Lets define a \textit{transversal XOR game} as an 
interactive protocol between a verifier and two entangled provers, Alice and 
Bob, that is specified by a function $g : \{0,1\}^m \times \{0,1\}^l 
\rightarrow \{0,1\}$ and a distribution $\pi$ on 
$\{0,1\}^m \times \{0,1\}^l$. The operation of the game is as 
follows.
\begin{enumerate}
\item 
The verifier generates $(z,r) \in \{0,1\}^m \times \{0,1\}^l$ 
according to distribution $\pi$. Then the verifier produces two 
shares of $z$, $s$ and $t$, by generating $s \in \{0,1\}^n$ 
uniformly and independently of $z$ and setting $t = s \oplus z$. The verifier sends 
$s$ to Alice and $t$ to Bob.
\item 
Alice and Bob produce bits $a$ and $b$, respectively, and send them 
to the verifier.
\item
The verifier accepts if and only if $a \oplus b = g(s \oplus t,r)$.
\end{enumerate}

The following is a slight generalization of a result 
in~\cite{LPSW06} and its proof appears in the Appendix~\ref{app:proof}.

\begin{theorem}\label{thm:lindenext}
Let $G$ be a transversal XOR game specified by $g : \{0,1\}^m \times 
\{0,1\}^l \rightarrow \{0,1\}$ and the distribution $\pi$. Then 
the optimal strategy that maximizes $\Pr[a \oplus b = 
g(s \oplus t, r)]$ does not use any entanglement and is of the 
following form. For some $u \in \{0,1\}^m$ and $\gamma \in \{0,1\}$ 
(that depend on $g$ and $\pi$), Alice responds with $a = (u \cdot s) 
\oplus \gamma$ and Bob responds with $b = u \cdot t$.
\end{theorem}

Now in our protocol $\cP$, the verifier on receiving $x$ could be
thought of as playing a transversal XOR game with the provers Alice
and Bob by letting $z \defeq e_i \oplus e_j \oplus e_k$, $r \defeq
\delta$ and $g: \{0,1\}^m
\times \{0,1\} \rightarrow \{0,1\}$ be such that $g(e_i \oplus
e_j \oplus e_k, \delta) \defeq f_x(i,j,k, \delta)$.  

Let $x \not\in L$. Now from Theorem~\ref{thm:lindenext} the
optimal strategy for the provers in which they are trying to maximize
the acceptance probability of the verifier $V$ would be as
follows. Alice and Bob ignore the entanglement and for some $u \in
\{0,1\}^{m}$ and $\gamma \in \{0,1\}$, Alice outputs $a = (u \cdot 
s) \oplus \gamma$ and Bob outputs $a = u \cdot t$.

Now it can be easily shown that $\Pr[V \mbox{ accepts } x] = \Pr[a
\oplus b = g(x,s\oplus t)] \leq 1/2 + \varepsilon $. Consider the
following PCP witness $w$. For all $j  \in \{1,2,\dots,m\}$, set $w_j
\defeq u_j \oplus \gamma$. Note that this  witness satisfies
\begin{eqnarray*}
w_i \oplus w_j \oplus w_k & = & u_i \oplus u_j \oplus u_k \oplus 
\gamma \\
& = &  u \cdot (e_i \oplus e_j \oplus e_k) 
\oplus \gamma \\
& = & u \cdot (s \oplus t) \oplus \gamma \\
& = & a \oplus b.
\end{eqnarray*}
Combining this with the fact that $f_x(i,j,k, \delta) = g(s \oplus t, \delta)$ 
enables us to conclude that
\[
\Pr[a \oplus b = g(x,s\oplus t)] = \Pr[w_i \oplus w_j \oplus w_k =
f_x(i,j,k, \delta)] \le \frac{1}{2} + \varepsilon. \]
The last inequality comes from the soundness property of the PCP
procedure $\Vpcp$. Thus $\Pr[V \mbox{ accepts } x]$ in the
protocol $\cP$ is at  most $\frac{1}{2} + \varepsilon$
and hence the soundness property is satisfied.


\appendix

\section{Proof of Theorem~\ref{thm:lindenext}}
\label{app:proof}
Our proof follows very much in the lines of the proof of Linden et 
al.~\cite{LPSW06}. Let Alice and Bob share a pure quantum state 
$\ket{\phi}$ between them. Let $Z,R$ be a pair of random variables 
jointly distributed according to $\pi$. Let $S,T \in \{0,1\}^n$ 
represent the random variables correspoding to the questions of 
verifier $V$ to Alice and Bob respectively. Let $A,B$ represent the 
random variables correspoding to the answers by Alice and Bob 
respectiely. Note that in our case $Z = S \xor T$ and $S$ is 
uniformly distributed and is independent of $(Z,R)$. It is well know 
that $\Pr[g(z, r) = A \xor B ~|~ (S,Z,R) = (s,z,r)]$ can be 
expressed as follows, $$ \Pr[g(z, r) = A \xor B ~|~ (S,Z,R) = 
(s,z,r)] = \frac{1}{2}(1 + (-1)^{g(z,r)} \bra{\phi} A_s \otimes B_{s 
\xor z} \ket{\phi})$$ , where $A_s, B_{s \xor z}$ are Hermitian 
operators with eigenvalues in $\{-1,1\}$, (which are also sometimes 
referred to as {\em observables}). Therefore we have,

\begin{eqnarray*}
\Pr[V \mbox{ accepts}] & =  & \sum_{s,z,r} \Pr[(S, Z,R)= (s,z,r)]
\Pr[g(z, r) = A \xor B ~|~ (S,Z,R) = (s, z, r)] \\
& = & \sum_{s,z,r} \Pr[(S, Z,R)= (s,z,r)] \cdot \frac{1}{2}(1 +
(-1)^{g(z,r)} \bra{\phi}  A_s \otimes B_{s \xor z}  \ket{\phi})\\
& =  & \frac{1}{2} + \sum_{s,z,r} \Pr[(S, Z,R)= (s,z,r)] \cdot 
\frac{1}{2}(-1)^{g(z,r)} \bra{\phi}  A_s \otimes B_{s \xor z}
\ket{\phi}) \\
& =  & \frac{1}{2} + \sum_{s,z} \frac{1}{2^n}  (\sum_r 
\Pr[(R,Z)=(r,z)] \cdot \frac{1}{2} (-1)^{g(z,r)} ) \bra{\phi}  A_s 
\otimes B_{s \xor z}  \ket{\phi}
\end{eqnarray*}

Now let,
\begin{eqnarray*}
& & \theta_z \defeq \sum_r \Pr[(R,Z)=(r,z)] \cdot \frac{1}{2} (-1)^{g(z,r)}  \\
& & \ket{\alpha} \defeq \frac{1}{\sqrt{2^n}} \sum_s (A_s \otimes
I)(\ket{\phi} \otimes \ket{s}) \\
& & \ket{\beta} \defeq \frac{1}{\sqrt{2^n}} \sum_t (I \otimes
B_t)(\ket{\phi} \otimes \ket{t})  \\
& & \Phi \defeq \sum_{s,z} \theta_z \ket{s}\bra{s \xor z}
\end{eqnarray*}
Note that as defined above $\ket{\alpha}, \ket{\beta}$ are unit 
vectors. Also $\Phi$ is Hermitian. Therefore from above we have,

\begin{eqnarray*}
\Pr[V \mbox{ accepts}] & =  &  \frac{1}{2} + \sum_{s,z} 
\frac{1}{2^n} \theta_z  \bra{\phi}  A_s \otimes B_{s \xor 
z}  \ket{\phi}
\\
& = & \frac{1}{2} + \bra{\alpha}(I \otimes \Phi)\ket{\beta} \\
&\leq& \frac{1}{2} + ||\bra{\alpha}||_2 ||(I \otimes
\Phi)||_{\infty}||\ket{\beta}||_2 \\ 
& =  & \frac{1}{2} + ||(I \otimes \Phi)||_{\infty} \\
& = &  \frac{1}{2} + ||\Phi||_{\infty}
\end{eqnarray*}
Above $||\Phi||_{\infty}$ represents the highest singular value of 
$\Phi$ and since $\Phi$ is Hermitian it means the highest modulus 
eigenvalue.

Now we show below that the eignevectors of $\Phi$ are precisely the 
{\em Hadamard} vectors $\ket{u} \defeq \sum_{v \in  \{0,1\}^n} 
(-1)^{u \cdot v}\ket{v} $ (for $u \in \{0,1\}^n$) with eigenvalues 
$\lambda_u = \sum_z (-1)^{u \cdot z} \theta_z $.  Consider,
\begin{eqnarray*}
\Phi \ket{u} & = & (\sum_{s,z} \theta_z \ket{s}\bra{s \xor 
z})
(\sum_{v \in  \{0,1\}^n} (-1)^{u \cdot v}\ket{v}) \\
& = &  \sum_{s,z} (-1)^{u \cdot (s \xor z)}  \theta_z \ket{s} \\
& = & (\sum_z (-1)^{u \cdot z}  \theta_z ) \sum_{s} (-1)^{u \cdot s} \ket{s} \\
& = &  \lambda_u \ket{u}
\end{eqnarray*}

Next we show that there exists a classical strategy by Alice and Bob 
such that $\Pr[V \mbox{ accepts}]  = \frac{1}{2} + 
||\Phi||_{\infty}$. Let $\ket{w}$ be the eigenvector of $\Phi$ 
corresponding to the highest modulus eigenvalue. Let $\gamma = 0$ if 
$\lambda_w \geq 0 $ and 1 otherwise. Now let Alice answer with $(w 
\cdot s) \xor \gamma$ to question $s$ and let Bob answer with  $(w 
\cdot t) $ to question $t$. Then we see that,
\begin{eqnarray*}
\Pr[V \mbox{ accepts}] & = &  \sum_{s,z,r} \Pr[(S, Z,R)= (s,z,r)]
\Pr[g(z, r) = (w \cdot s)  \xor \gamma \xor  (w \cdot (s \xor z))] \\
& = & \sum_{s,z,r} \Pr[(S, Z,R)= (s,z,r)]
\Pr[g(z, r) = (w \cdot z) \xor \gamma ] \\
& = & \sum_{s,z,r} \Pr[(S, Z,R)= (s,z,r)]
( \frac{1}{2} + \frac{1}{2}(-1)^{g(z, r) +  (w \cdot z) + \gamma}) \\
& = & \frac{1}{2} + \sum_{s,z,r} \Pr[(S, Z,R)= (s,z,r)] \cdot 
\frac{1}{2} \cdot (-1)^{g(z, r) +
(w \cdot z) + \gamma}  \\
& = & \frac{1}{2} + \sum_{z,r} \Pr[(Z,R)= (z,r)] \cdot \frac{1}{2}
\cdot (-1)^{g(z, r) + (w \cdot z) + \gamma}  \\
& = & \frac{1}{2} + \sum_z (\sum_r \frac{1}{2} \cdot \Pr[(Z,R)= (z,r)]
\cdot (-1)^{g(z,r)}) (-1)^{(w \cdot z) + \gamma} \\
& = & \frac{1}{2} + \sum_z \theta_z \cdot (-1)^{(w
\cdot z) + \gamma} \\
& = & \frac{1}{2} + |\lambda_w| = \frac{1}{2} + ||\Phi||_\infty
\end{eqnarray*}

\qed


\begin{thebibliography}{10}



\bibitem{BabaiF+91}
L.~Babai, L.~Fortnow, and C.~Lund.
\newblock Non-deterministic exponential time has two-prover interactive
  protocols.
\newblock {\em Computational Complexity}, 1(1):3--40, 1991.




\bibitem{Ben-OrG+88}
M.~Ben-Or, S.~Goldwasser, J.~Kilian, and A.~Wigderson.
\newblock Multi-prover interactive proofs: how to remove intractability
  assumptions.
\newblock In {\em Proceedings of the Twentieth Annual ACM Symposium on Theory
  of Computing}, pages 113--131, 1988.



\bibitem{CleveH+04a}
R.~Cleve, P.~H\o yer, B.~Toner, J.~Watrous.
\newblock Consequences and limits of nonlocal strategies.
\newblock In {\em Proceedings of the 19th IEEE Conference on Computational
Complexity}, pages 236--249, 2004.



\bibitem{DolevD+00}
D.~Dolev, C.~Dwork and M.~Naor. 
\newblock Non-malleable cryptography. 
\newblock {\em SIAM Journal on Computing} 30(2):391--437, 2000.
 
\bibitem{Feige91}
U.~Feige.
\newblock On the success probability of two provers in one-round proof systems.
\newblock In {\em Proceedings of the Sixth Annual Conference on Structure in
  Complexity Theory}, pages 116--123, 1991.

\bibitem{FeigeL92}
U.~Feige and L.~Lov\'asz.
\newblock Two-prover one-round proof systems: their power and their problems.
\newblock In {\em Proceedings of the Twenty-Fourth Annual ACM Symposium on
  Theory of Computing}, pages 733--744, 1992.



\bibitem{FortnowR+94}
L.~Fortnow, J.~Rompel, and M.~Sipser.
\newblock On the power of multi-prover interactive protocols.
\newblock {\em Theoretical Computer Science}, 134:545--557, 1994.

\bibitem{Hastad01}
J.~H{\aa}stad.
\newblock Some optimal inapproximability results.
\newblock {\em Journal of the ACM}, 48(4):798--859, 2001.

\bibitem{KempeVidick06}
J.~Kempe and T.~Vidick.
\newblock On the power of entangled quantum provers.
\newblock arXiv:quant-ph/0612063, 2006.

\bibitem{KKMTV07}
J.~Kempe, H.~Kobayashi, K.~Matsumoto, B.~Toner, and T.~Vidick.
\newblock On the power of entangled provers: immunizing games against
\newblock entanglement.
\newblock Manuscript, 2007.

\bibitem{KitaevW00}
A.~Kitaev and J.~Watrous.
\newblock Parallelization, amplification, and exponential time simulation of 
\newblock quantum interactive proof systems.
\newblock In {\em Proceedings of the Thirty-Second Annual ACM Symposium on
  Theory of Computing}, pages 608--617, 2000.

\bibitem{KobayashiM03}
H.~Kobayashi and K.~Matsumoto.
\newblock Quantum multi-prover interactive proof systems with limited 
\newblock prior entanglement.
\newblock {\em Journal of Computer and System Sciences}, 66(3):429--450, 2003.


\bibitem{LPSW06}
N.~Linden, S.~Popescu, A.~J.~Short, and A.~Winter.
\newblock No quantum advantage for nonlocal computation.
\newblock arXiv:quant-ph/0610097, 2006.


\bibitem{Watrous99}
J.~Watrous.
\newblock PSPACE has constant-round quantum interactive proof systems.
\newblock in {\em Proceedings of the Fourtieth Annual Symposium on Foundations
of Computer Science}, pages 112--119, 1999.

\bibitem{Wehner06}
S.~Wehner.
\newblock Entanglement in interactive proof systems with binary answers.
\newblock In {\em Proceedings of STACS 2006}, pages 162--171, 2006.

\end{thebibliography}
\end{document}